\documentclass[preprints,article,accept,oneauthor,pdftex]{Definitions/mdpi} 

\firstpage{1} 
\pubvolume{xx}
\issuenum{x}
\articlenumber{x}
\pubyear{2021}
\copyrightyear{2021}
\history{Received: date; Accepted: date; Published: date}

\Title{Our Peculiar Motion Inferred from Number Counts of Mid Infra Red AGNs and 
the Discordance Seen with the Cosmological Principle}

\Author{Ashok K. Singal}

\address{%
\quad Astronomy and Astrophysics Division, Physical Research Laboratory,
Navrangpura, Ahmedabad - 380 009, India; ashokkumar.singal@gmail.com
}

\corres{Correspondence: ashokkumar.singal@gmail.com; Tel.: +91-9427-631-833}

\abstract{The dipole anisotropy in the Cosmic Microwave Background Radiation (CMBR) has given a peculiar velocity vector 370 km s$^{-1}$ along $l=264^\circ,b=48^\circ$. However, some other dipoles, for instance, from the number counts, sky brightness or redshift distributions in large samples of distant Active Galactic Nuclei (AGNs), have yielded values of the peculiar velocity many times larger than that from the CMBR, though surprisingly, in all cases the directions agreed with the CMBR dipole. Here we determine our peculiar motion from a sample of ~0.28 million AGNs, selected from the Mid Infra Red Active Galactic Nuclei (MIRAGN) sample comprising more than a million sources. From this, we find a peculiar velocity, which is more than four times the CMBR value, although the direction seems to be within $\sim 2\sigma$ of the CMBR dipole. A  genuine value of the solar peculiar velocity should be the same irrespective of the data or the technique employed to estimate it. Therefore, such discordant dipole amplitudes, might mean  that the  explanation for these dipoles, including that of the CMBR, might in fact be something else. But, the observed fact that the direction in all cases, is the same, though obtained from completely independent surveys using different instruments and techniques, by different sets of people employing different computing routines, might nonetheless indicate that these dipoles are not merely due to some systematics, otherwise why would they all be pointing along the same direction. It might instead suggest a preferred direction in the Universe, implying a genuine anisotropy, which would violate the Cosmological Principle, the core of the modern cosmology.}


\begin{document}



\section{Introduction}
According to the Cosmological Principle, the Universe, when seen on a sufficiently large scale (beyond a few hundred Mpc), should appear isotropic, without any preferred directions, to a co-moving observer in the expanding Universe. Such an observer is at rest with respect to the Universe at large and the angular distribution of sources in sky should appear statistically to be similar in all directions. However, if relative to the co-moving coordinates the observer has a motion, called a peculiar motion, then because of the Doppler boosting as well as   aberration effects, the observer will find the sky brightness as well as the number counts of distant extragalactic objects to manifest a dipole anisotropy, proportional to the peculiar velocity of the observer. The Cosmic Microwave Background Radiation (CMBR), shows such a dipole anisotropy that, when ascribed to the peculiar motion of the Solar system, yields for the peculiar velocity a value $370$ km s$^{-1}$ along right ascension (RA) $=168^{\circ}$, declination (Dec) $=-7^{\circ}$, or in galactic coordinates, $ l=264^{\circ}, b=48^{\circ}$ \cite{1,2,3}. 
\begin{figure*}
\centering
\includegraphics[width=\linewidth]{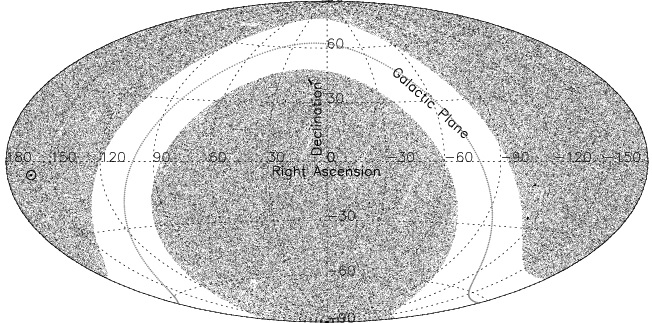}
\caption{The sky distribution of ~0.28 million AGNs of our MIRAGN sample (see text), in the Hammer-Aitoff equal-area projection map, plotted in equatorial coordinates with right ascension (RA) from $-180^\circ$ to $180^\circ$ and declination (Dec) from $-90^\circ$ to $90^\circ$, shows the sources to be spread quite uniformly across the sky, except for the gap seen in the $\pm 15^\circ$ wide strip about the galactic plane, a zone of exclusion.  The sky position of the CMBR pole is indicated by $\odot$.
}
\end{figure*}

On the other hand, our peculiar velocity has also been determined in recent years from the anisotropy observed in the sky distribution of large samples of discrete radio sources. The NRAO  VLA Sky Survey (NVSS), comprising 1.8 million radio sources \cite{12}, showed a statistically significant dipole asymmetry corresponding to a velocity $\sim 4$ times the CMBR value \cite{4}, something that was not only unexpected, but appeared initially almost preposterous, however, confirmed subsequently by many independent groups \cite{5,6,7,8}. Further, in the TIFR GMRT Sky Survey (TGSS) \cite{14}, comprising 0.62 million sources \cite{13}, a very significant ($>10\sigma$) dipole anisotropy, amounting to a velocity $\sim 10$ times the CMBR value, was detected \cite{8,9}. However, equally surprising, the direction of motion in both cases has turned out to be along the CMBR dipole. Recently, a homogeneously selected DR12Q sample of 103245 distant quasars has shown a redshift dipole along the CMBR dipole direction, implying a velocity $\sim 6.5$ times though in a direction directly opposite to, but nonetheless parallel to, the CMBR dipole \cite{10}. A more recent determination of the peculiar motion from a sample of quasars derived from the Wide-field Infrared Survey Explorer (WISE), has shown an amplitude over twice as large the CMBR value \cite{11}.  Now a genuine solar peculiar velocity cannot vary from one set of measurements to another and such discordant dipoles could imply that the explanation for the genesis of these dipoles, including that of the CMBR, might lie elsewhere. At the same time a common direction for all these dipoles, determined from completely independent surveys by different groups, using independent computational routines, does indicate that the differences in  the dipoles are not merely random fluctuations or due to some systematics in data or procedures, otherwise their directions too would be different. Instead, it might  suggest a preferred direction in the Universe implying a genuine anisotropy, which would violate the Cosmological Principle, the core of the modern cosmology. Because of the huge impact on the cosmological models any genuine variations in the dipole magnitudes may impart, further independent determinations of the dipole vectors are warranted. 

Here we determine our peculiar motion from a sample of ~0.28 million AGNs (Figure 1), selected from the Mid Infra Red Active Galactic Nuclei (MIRAGN) sample comprising more than a million sources \cite{17}. A preliminary account of these results was presented in proceedings of the 1st Electronic Conference on Universe \cite{si21}, where the genesis of a dipole anisotropy in the number density, owing to the observer's peculiar motion, in an otherwise isotropic distribution of sources in the sky, and how one could estimate the peculiar velocity vector from the observed dipole anisotropy and the sample selected by us for this purpose, were discussed. Here we pursue and explore these findings in a greater detail, discussing any possible pitfalls in the sample selection and the techniques employed. Further, we exploit  another alternate method to extract the dipole, and do a comparison of our derived results vis-\`a-vis the peculiar velocity information previously available in the literature, and point out their cosmological implications.
\begin{figure*}
\centering
\includegraphics[width=\linewidth]{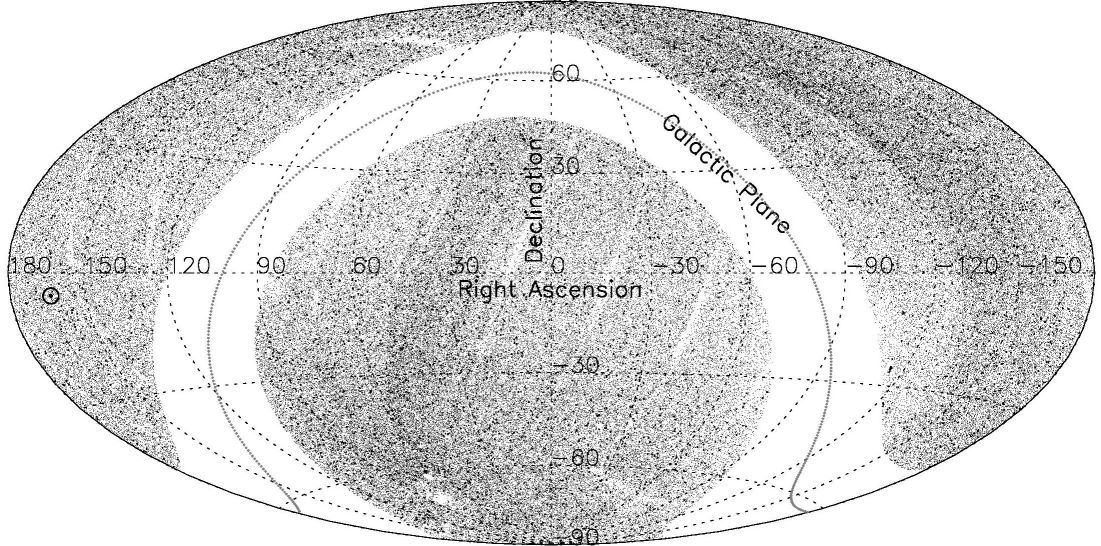}
\caption{The sky distribution of ~0.27 million MIRAGNs in the range  $15.5<W1<16$, in the Hammer-Aitoff equal-area projection map, plotted in equatorial coordinates with RA from $-180^\circ$ to $180^\circ$ and Dec from $-90^\circ$ to $90^\circ$. The sky position of the CMBR pole is indicated by $\odot$. The sky distribution of these sources, which are at fainter levels than those in Figure 1, shows a non-uniformity that is quite evident in the figure and seems to the result of a deeper coverage in certain regions of sky.}
\end{figure*}
\section{Dipole Vector Due to the Motion of the Observer}
Due to the assumed isotropy of the Universe -- \`a la cosmological principle -- an observer stationary with respect to the comoving coordinates of the cosmic fluid, should find the average number densities of distant AGNs as well as their flux densities to be distributed uniformly over the sky. 
However, an observer moving with a velocity $v$ relative to the cosmic fluid, will find a source along an angle $\theta$ with respect to the direction of motion, to appear brighter due to Doppler boosting by a factor $\propto \delta^{(1+\alpha)}$ \cite{15}, where $\delta=1+(v/c)\cos\theta$ is the Doppler factor and $\alpha$ is the spectral index defined by $S \propto \nu^{-\alpha}$, with $S$ being the flux density of the source at the frequency $\nu$ of the  observing instrument. 
Here we have used the non-relativistic formula for the Doppler factor as all previous observations indicate that $v\ll c$. As the integral source counts of the extragalactic source population usually follow a power law, $N(>S)\propto S^{-x}$ \cite{15}, the number of sources observed by a telescope of a given sensitivity will be higher by a factor  $\propto \delta^{x(1+\alpha)}$, due to the Doppler boosting \cite{15}. 
Additionally, due to the aberration of light, the apparent position of a source will shift toward the direction of motion by a value, $(v/c)\sin\theta$, thereby changing the number density by another factor  $\propto \delta^{2}$ \cite{15}. 
Thus, as a combined effect of Doppler boosting and aberration, the observed number counts will vary with direction as $\propto \delta^{2+x(1+\alpha)}$, which, for $v\ll c$, can be expressed as a dipole anisotropy, $1+{\cal D}\cos\theta$ \cite{4,15,16}, with an amplitude 

\begin{equation}
\label{eq:1}
{\cal D}=\left[2+{x(1+\alpha)}\right]\frac{v}{c}\;.
\end{equation}

Let  $ \bf{\hat{r}}_i$ be the position vector of $i^{th}$ source, then a stationary observer,  due to the assumed isotropy of the Universe, should find $\Sigma \bf{\hat{r}}_i=0$. However, for a moving observer, $\Sigma \bf{\hat{r}}_i$ will yield a net vector along the direction of motion \cite{16}. Then the peculiar speed $v$ of the observer could be obtained from   \cite{4,9}

\begin{equation}
\label{eq:2}
{\cal D}=\frac{3}{2}\frac{\Sigma \cos \theta_i}{\Sigma |\cos \theta_i|}= \frac{3\Sigma \cos \theta_i}{N}\:,
\end{equation}
where $\theta_i$ is the polar angle of the $i^{th}$ source with respect to the dipole direction of motion \cite{16} and $N$ is the total number of sources in the sample. 

Thus, exploiting the angular positions of extragalactic sources in a large survey that covers the whole sky and is complete in the sense that it comprises all sources above a certain flux-density limit, we can determine our peculiar motion. It should be noted that exclusions of strips like the galactic plane, $|b| < 15^\circ$ (Figure 1), which affect the forward and backward measurements identically, do not have systematic effects on the determined direction of the peculiar motion \cite{4,15}. 
\begin{figure*}
\centering
\includegraphics[width=\linewidth]{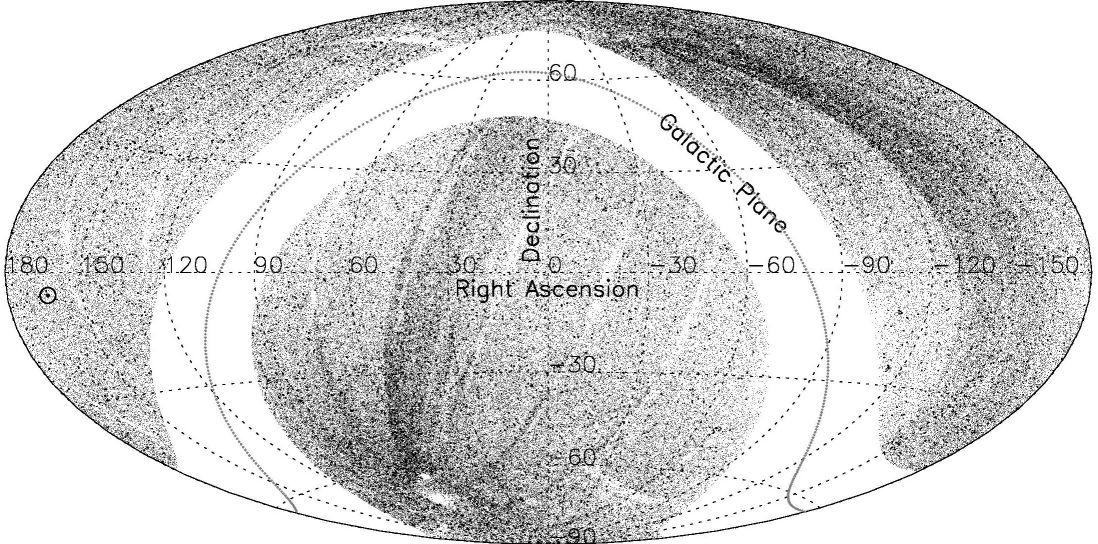}
\caption{The sky distribution of ~0.32 million MIRAGNs in the range $16<W1<17$, in the Hammer-Aitoff equal-area projection map, plotted in equatorial coordinates with RA from $-180^\circ$ to $180^\circ$ and Dec from $-90^\circ$ to $90^\circ$. The sky position of the CMBR pole is indicated by $\odot$. The non-uniformity in the sky distribution of these sources, which are at still fainter levels than those in Figure 2, is even higher, and seems to arise from a deeper coverage in certain regions of sky.}
\end{figure*}

\section{Our Sample of MIRAGNs}
The sample of AGNs used in this study is selected from a publicly made available larger all-sky sample of 1.4 million active galactic nuclei (AGNs) \cite{17}, in turn derived from the Wide-field Infrared Survey Explorer final catalog release (AllWISE), that incorporates data from the WISE Full Cryogenic, 3-Band Cryo, and NEOWISE Post-Cryo survey \cite{18,19}. The WISE survey is an all-sky mid-IR survey at 3.4, 4.6, 12, and 22 $\mu$m (W1, W2, W3, and W4) with angular resolutions 6.1, 6.4, 6.5 and 12 arcsec, respectively. AllWISE comprises data for almost 748 million objects, out of these about 1.4 million objects met a two-color infrared photometric selection criteria for AGNs, that formed the original MIRAGN sample \cite{17}. 	

For our purpose we have restricted the MIRAGN sample to an upper limit of magnitude, W1$<15.0$, mainly because of a differential increase in the number density for weaker sources in various regions of the sky, 
due to deeper WISE coverage. However, due to the completeness of the basic survey at strong source levels, the number density distribution in the sky at low infrared magnitudes remains unaffected as a deeper coverage adds only fainter sources, which are at higher infrared magnitudes. From a detailed examination of the original MIRAGN sample data in small-range magnitude slices at different W1 levels, we find that from W1$\approx 15.5$ onward,  there is a non-uniform distribution in sky that increases rapidly at higher magnitudes, i. e., for weaker sources. Figures (2) and (3)) show the increasing non-uniformity for magnitude levels W1$ > 15.5$. Accordingly, we have chosen W1=15.0 to be our upper magnitude limit. On the lower side, we have restricted our sample to W1$>12.0$. This is only to minimize the effects of the local bulk flows which will affect sources at low redshifts, $z<0.05$, corresponding to W1<12. In any case, the number of sources for W1$<12$ is relatively very small and their inclusion or exclusion hardly affects the results. Further, we have also excluded all sources in the galactic plane with $|b|<15^\circ$ to avoid contamination by galactic sources \cite{17}. Our final sample then comprises 279139 AGNs. Figures (1) shows the sky distribution of sources in our sample. The distribution seems to be quite uniform across the sky, except for the gap seen in the $\pm 15^\circ$ wide strip about the galactic plane.
\begin{figure*}[t]
\centering
\includegraphics[width=\linewidth]{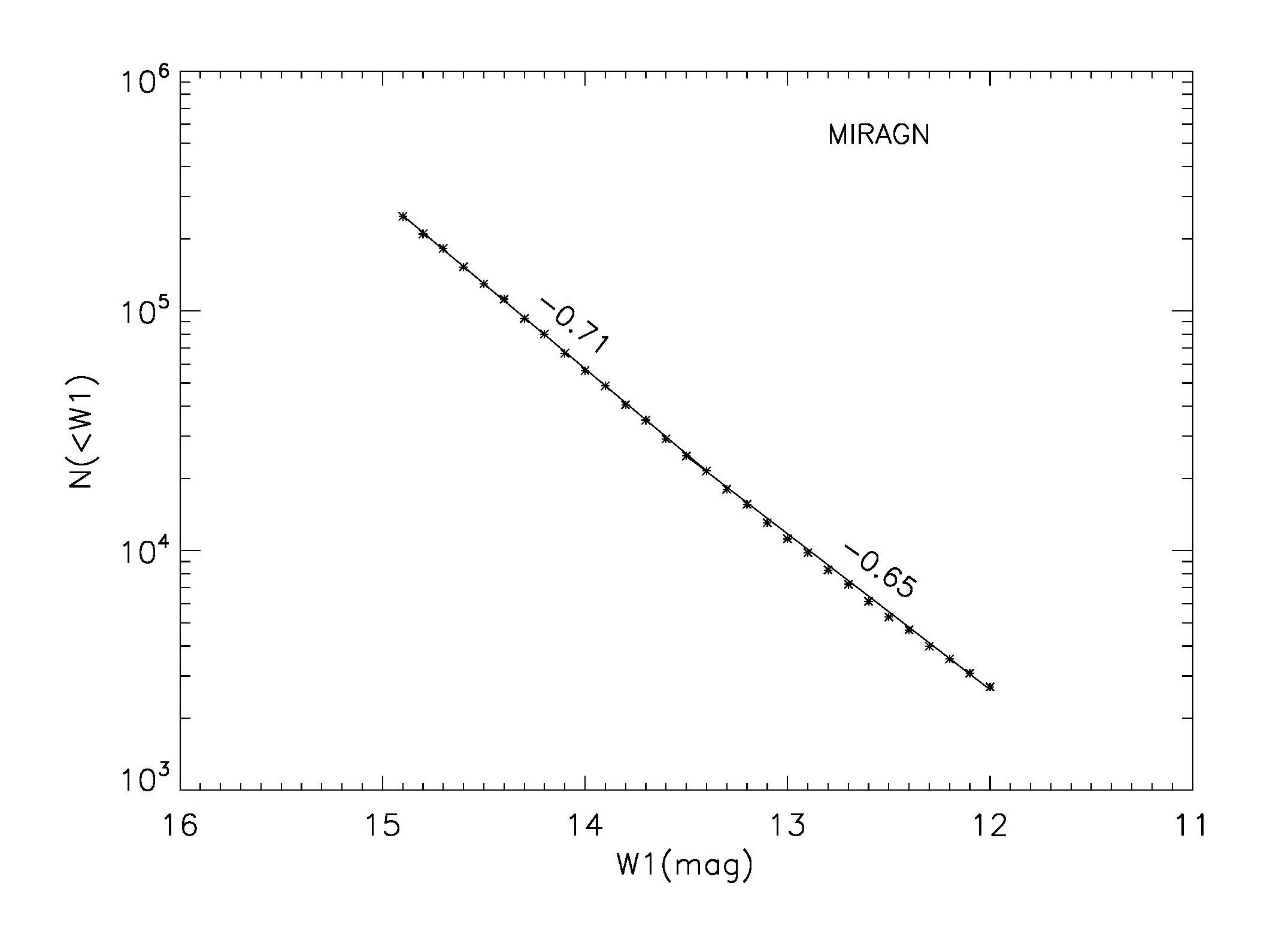}
\caption{A plot of the integrated source counts N(<W1) against W1, for our MIRAGN sample, showing the power law behaviour of the source counts.}
\end{figure*}
 
\section{Results}
We determine the dipole using two alternate methods from the same data set of angular sky positions of MIRAGNs in our sample. In the computation of the dipole, the weight that a source gets in the two methods is different, depending upon its angular position in the sky.
\subsection{Dipole Vector Determined Directly from the Sum of Position Vectors of AGNS}
Figure 1 shows the sky distribution of all ~0.28 million MIRAGN AGNs in our sample, in the Hammer-Aitoff equal-area projection map, plotted in equatorial coordinates. The source distribution looks quite uniform over the sky, except for the gap in the galactic plane band, where we have removed a $\pm 15^\circ$ band about the galactic plane.  As mentioned earlier, such exclusions, which affect the forward and backward measurements identically, do not have systematic effects on the results, as far as the direction of the dipole is concerned. However, the estimated dipole amplitude $\cal D$ might be affected by the $\pm 15^\circ$ gap in the sky coverage about the galactic plane. Let $k_1$ be the factor, of the order of unity, to be determined numerically, by which the dipole amplitude gets overestimated due to the presence of gaps in the sky coverage, and with which $\cal D$ should be divided while computing $v$. 

Before proceeding with the actual source sample we made Monte-Carlo simulations with source distributions similar to that in our sample. In each simulation, for the magnitude (W1) distributions we took the actual values as in our sample, however, the sky position was allotted randomly for each one of the 0.28 million sources. On this were superimposed Doppler boosting and aberration effects of an assumed peculiar motion of the observer, choosing a different velocity vector for each simulation. The resultant artificial sky was then used to recover back the velocity vector and compared with the value actually used in that particular simulation. This not only verified our procedure and the computation routine, but also allowed us to make an estimate of errors in the dipole direction from the spread observed in 500 independent simulations. The correction factor, determined from another set of 500 Monte-Carlo simulations run without any gap about the galactic plane, turned out to be only  $k_1\approx 1.01$

The results for the dipole, determined from the anisotropy in number counts in our sample of about 0.28 million MIRAGNs in the $12<$W1$<15$ range, are given in Table 1, where the 1st column gives the magnitude range, 2nd column gives the number of sources, 3rd and 4th columns give the direction of the dipole in terms of Right Ascension and Declination, 5th column gives $\cal D$, the dipole magnitude, and the last column gives the value of the speed estimated from $\cal D$.

We find a peculiar speed $1.7 \pm 0.2 \times 10^3$ km s$^{-1}$, which is more than four times the peculiar speed estimated from the CMBR dipole. In order to ensure that this excess in speed is not due to a skew distribution of sources belonging to a particular magnitude, we divided our sample into three magnitude bins, with approximately 0.1 million sources in each bin. The results for the three bins are also presented in Table I. Here, N is the total number of sources in the corresponding W1 bin, RA and Dec give the dipole direction in the sky, $\cal D$ is the dipole value computed from ${\cal D}={3\Sigma \cos \theta_i}/(2\Sigma |\cos \theta_i)]$, with error $\Delta {\cal D} =\sqrt{3/(2\Sigma |\cos \theta_i|)}$ \cite{4,9,16}. Then the peculiar speed of the observer, or rather of the Solar system, is computed from ${\cal D}=[2+x(1+\alpha)](v/c)$. In order to determine $x$ we have made a plot of the integrated source counts N(<W1) against W1, for our MIRAGN sample, in Figure~4, which shows a power law behaviour of the integrated source counts, with a slope that varies between 0.71 and 0.65, with a mean value of 0.68. From this we estimate the index of integral source counts in our sample to be $x=2.5 \times 0.68=1.7$,  consistent with the value quoted in the literature  \cite{11}. For the spectral index, we have taken $\alpha\approx 1$, a value quoted in the literature \cite{11}, for the extragalactic population of AGNs. The peculiar speed, accordingly, is given by, $v\approx c D /(5.4 k_1)\approx 5.5 \times 10^4 {\cal D}$ km/s. 
\begin{table*}[t]
\caption{Peculiar velocity estimated from the dipole asymmetry in number counts, for $|{\rm b}|>15^\circ$.}
\hskip4pc\vbox{\columnwidth=33pc
\begin{tabular}{ccccccccccccccccc}
 \hline
(1)&(2)&&(3)&&(4)&&(5)&&(6)\\Magnitude Range & $N$ &&  RA && Dec && ${\cal D}$   &&  $v$\\
  && & ($^{\circ}$)& & ($^{\circ}$) && ($10^{-2}$) && ($10^{3}$ km s$^{-1}$) \\ \hline
$15.0>W1\geq 12.0$ & 279139 && $148 \pm 19$ && $23\pm 17$ && $3.0 \pm 0.4$ && $1.7 \pm 0.2$ \\
$15.0>W1\geq 14.7$ & 102822 && $157 \pm 20$ && $23 \pm 18$ && $4.1\pm 0.6$ && $2.3\pm 0.3$ \\
$14.7>W1\geq 14.3$ & 086035  && $132 \pm 21$ && $32 \pm 19$ && $2.9\pm 0.6$ && $1.6\pm 0.3$ \\
$14.3>W1\geq 12.0$ & 090282  && $143 \pm 21$ && $11 \pm 19$ && $2.1\pm 0.6$ && $1.2\pm 0.3$ \\
\hline
\end{tabular}
}
\end{table*}

In Table 1, we have also listed the direction of the dipole,  along with the 
estimated errors, as determined in each case. The direction of the velocity vector 
(with our best estimate from Table-1), is given by the pole at RA$= 148^\circ\pm 19^\circ$, Dec= $23^\circ\pm 17^\circ$, within $\stackrel{<}{_{\sim}} 2\sigma$ of the CMBR pole at RA$= 168^\circ$, Dec$= -7^\circ$ (in galactic coordinates, the MIRAGN pole lies at $l = 209^\circ, b = 49^\circ$ while the CMBR pole is at $l = 264^\circ, b = 48^\circ$, with the CMBR pole positions being $~55 \cos 49\approx 38^\circ$ away). However the estimates of $v (=1.7 \pm 0.2 \times 10^3$ km s$^{-1}$) appear much higher than the CMBR value (370 km s$^{-1}$) by a factor $>4$ at a statistically significant ($\stackrel{>}{_{\sim}} 5\sigma$) level, which however, is in agreement with the dipole derived from the NVSS radio source data \cite{4,5,6,7,8}. 

Although we have restricted our sample to W1$>12.0$ to minimize the effect of local bulk flows, still, in order to estimate the influence on our results of any local clustering, like the Virgo super-cluster, we determined dipole vectors by excluding sources at low super-galactic latitudes, progressively in steps of five degrees. 
Table 2 shows the results where the 1st column gives the $|SGB|$ limit, 2nd column gives the number of sources, 3rd and 4th columns give the direction of the dipole in terms of Right Ascension and Declination, 5th column gives $\cal D$, the dipole magnitude,
and the last column gives the value of the speed estimated from ${\cal D}$.
\begin{table*}[t]
\begin{center}
\caption{Dipole estimates for various $|{\rm SGB}|$ limits}
\hskip4pc\vbox{\columnwidth=33pc
\begin{tabular}{lcccccccccccccc}
 \hline
 \;\;\;\;\;\;\;(1)&(2)&&(3)&&(4)&&(5)&&(6)\\ 
$|{\rm SGB}|$ limit & $N$ &&  RA && Dec && ${\cal D}$   &&  $v$\\
  && & ($^{\circ}$)& & ($^{\circ}$) && ($10^{-2}$) && ($10^{3}$ km s$^{-1}$) \\ 
\hline
$|{\rm SGB}|\geq 0$ & 279139 && $148 \pm 17$ && $23 \pm 17$ && $3.0\pm 0.4$ && $1.7\pm0.2$\\
$|{\rm SGB}|\geq 5$ & 251281 && $145 \pm 18$ && $21 \pm 18$ && $3.1\pm 0.4$ && $1.7\pm0.2$\\
$|{\rm SGB}|\geq 10$ & 223230 && $142 \pm 18$ && $20 \pm 19$ && $3.4\pm 0.4$ && $1.9\pm0.2$\\
$|{\rm SGB}|\geq 15$ & 195664 && $141 \pm 19$ && $12 \pm 20$ && $3.7\pm 0.4$ && $2.0\pm0.2$\\
$|{\rm SGB}|\geq 20$ & 168826 && $136 \pm 20$ && $09 \pm 20$ && $3.7\pm 0.4$ && $2.0\pm0.2$\\
 \hline
\end{tabular}
}
\end{center}
\end{table*}

From a comparison of the results in these cases ($|SGB|>0^\circ,5^\circ,10^\circ,15^\circ,20^\circ$; Table 2),  no unusually large variations, beyond the statistical uncertainties, were seen in the computed dipole vectors.

\subsection{Dipole determined from the hemisphere method} 
We can also determine the dipole employing an alternate method known as the hemisphere method.
Though this method appears to be simpler in nature, but here, unlike the previous dipole vector method, one does not directly get the direction of the dipole, which one has to assume or determine in some other way. Suppose we know the pole direction in sky, then 
using the great circle at $90^\circ$ from the pole direction, we divide the sky in two equal hemispheres, ${\cal S}_1$ and ${\cal S}_2$; ${\cal S}_1$ containing the assumed pole and ${\cal S}_2$, the opposite hemisphere, containing the antipole. 

As we saw earlier, a motion of the observer toward the pole will result in a dipole anisotropy, $n_0(1+{\cal D}\cos\theta)$, in the number counts of sources, where $n_0$ is the mean number density (per steradian) in sky and $\theta$ is the angle measured from the pole. 
Then the number of sources in the hemisphere ${\cal S}_1$ should be larger than the number of sources in ${\cal S}_2$.

If ${N}_1(\theta)$ be the cumulative number of sources in the sky zone between the great circle at $\theta=90^\circ$ and a parallel circle at angle $\theta$ in ${\cal S}_1$, while 
${N}_2(\theta)$ is the cumulative number of sources in the symmetrically placed, corresponding zone in the opposite hemisphere ${\cal S}_2$, i.e., ${N}_2(\theta)$ is the cumulative number of sources lying between $\theta=90^\circ$ and $\pi -\theta$. Then we obtain the dipole component ${\cal D}_{\theta}$ along $\theta$, from \cite{9}

\begin{equation}
\label{eq:8}
{\cal D}_{\theta}={\cal D}\cos\theta=\frac {{N}_1(\theta) - {N}_2(\theta)}{[{N}_1(\theta) + {N}_2(\theta)]/2}.
\end{equation}
The $1\sigma$ uncertainty in ${\cal D}_{\theta}$ is $2/\sqrt{{N}_1(\theta) + {N}_2(\theta)}$. 

Amplitude of the dipole ${\cal D}$ could then be determined by counting  
${N}1$ and ${N}2$ for the complete hemispheres, i.e., from  $\theta=90^\circ$ to $\theta=0^\circ$ in ${\cal S}_1$ and  $\theta=90^\circ$ to $\theta=180^\circ$ in ${\cal S}_2$, to get 

\begin{equation}
\label{eq:9}
{\cal D}= {{\cal D}_{\rm o}}= \frac {{N}_1 - {N}_2}{{N}/2}
=\frac {{N}_1 - {N}_2}{2\pi n_0}
\,,
\end{equation}
where ${N}={N}_1 + {N}_2=4\pi n_0$ is the total number of sources integrated over all directions in our sample, with the $1\sigma$ uncertainty in ${\cal D}$ being $2/\sqrt{{N}}$. 
\begin{figure*}
\centering
\includegraphics[width=\linewidth]{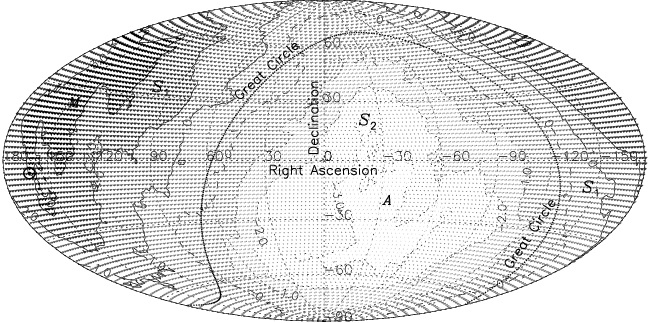}
\caption{A contour map of the dipole amplitudes, estimated for various directions in the sky. The horizontal and vertical axes denote RA and Dec in degrees. The true pole direction is expected to be closer to the higher contour values, indicated by darker grey regions, surrounded by continuous lines, while the  the true antipole should lie closer to the lower contour values, indicated by lighter grey regions, surrounded by dotted lines. 
The symbol $\odot$ indicates the CMBR pole position, while $M$ indicates the pole position for our MIRAGN sample, as determined by the Dipole Vector method (Table 1). Symbol $A$ indicates the corresponding antipole position. The dashed curve, representing the zero amplitude of the dipole, criss-crosses the Great Circle, at $90^\circ$ from the poles at $M$ and $A$, dividing the sky in two equal hemispheres, ${\cal S}_1$ and ${\cal S}_2$.} 
\end{figure*}

To determine the direction of the dipole, we have employed a 'brute force method', by  dividing the sky into cells of $2^\circ \times 2^\circ$ with minimal overlap, creating a grid of $10360$ pixels covering the sky. 
Then assuming the centre of each of these 10360 pixels in turn to be the pole direction, we computed the dipole magnitude from the sky positions of our 0.28 million MIRAGNs. This on the average yields only a projection of the peculiar velocity in the direction of each pixel, peaking in the pixel that lies closest to the true dipole direction. 

\begin{figure*}[h]
\centering
\includegraphics[width=12cm]{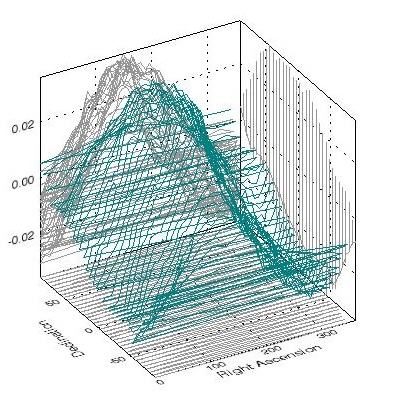}
\caption{A plot of dipole amplitudes (teal colour) estimated for various directions in the sky. The horizontal axes denote Right Ascension (RA) and Declination (Dec) in degrees. The positions (RA and Dec) of the peak values could be read from the 2-d projections along each axis, shown in light grey. Due to fluctuations in individual values it is not possible to determine a unique peak, whose location would have unambiguously determined the direction of our peculiar motion in the sky.} 
\end{figure*}
Figure~5 shows a contour map of the dipole amplitudes, estimated this way for various pixels on the sky. We should expect the true pole direction to be closer to the higher contour values, indicated by darker grey regions, surrounded by continuous lines, while the  the true antipole should lie closer to the lower contour values, indicated by lighter grey regions, surrounded by dotted lines. 
The pole position determined by the Dipole Vector method (Table 1) for our MIRAGN sample and denoted by $M$ here, lies almost in the middle of the darkest region, while the corresponding antipole position, indicated by the Symbol $A$, lies in the lighter-most region. The CMBR pole position, indicated by the symbol $\odot$, is not far from the peak of the dipole amplitude, indicated by the highest contour. The dashed curve, representing the zero amplitude of the dipole, criss-crosses the Great Circle, at $90^\circ$ from the poles at $M$ and $A$, dividing the sky in two equal hemispheres, ${\cal S}_1$ and ${\cal S}_2$. 

A 3-d representation of the dipole amplitude distribution across the sky is shown in Figure~6. 
Although an overall trend, in agreement with the dipole direction from Table 1, is seen in Figure~6, and where in principle, the location of the peak value for the dipole amplitude should yield the true direction of the dipole.
However, due to fluctuations in individual values it is not possible to determine the location of an unambiguous peak from Figure~6. At the same time, for an actual dipole, one would expect on the average a $\cos\theta$ dependence of the determined dipole magnitudes with respect to the true pole. To exploit this expectation, we have written a 3-d COSFIT routine, where for each of the $n=10360$ pixel positions in turn, we applied  a 3-d $\cos\theta$ fit to the remaining $n-1$ dipole values in the  pixels around it on the sky, and determined the resulting amplitude  as well as a chi-square value, for each of these $n$ fits. 

Figure~7 shows the output of our 3-d COSFIT routine. The process converged rapidly to show a clear unique peak at RA$=148^{\circ}$, Dec$=22^{\circ}$, with a maximum value of ${\cal D}=3.2\times 10^{-2}$, accompanied by an ideal minimum value of 1.0 for the reduced chi-square $\chi^2_\nu$,  occurring for the same pixel position as the peak. Figure~7(a) shows the 3-d plot of the peak while a clear minimum in the reduced $\chi ^2$ value is seen in Figure~7(b), both occurring at RA$=148^{\circ}$, Dec$=22^{\circ}$. In order to make sure that nothing further is amiss in our procedure, we also tried a finer grid with $1{^\circ} \times 1{^\circ}$, with more than $41000$ pixels, or even a coarser one with a pixel size of $5{^\circ} \times 5{^\circ}$,
but it made no perceptible difference in our final results. 

To test our 3-d COSFIT procedure, we made Monte-Carlo simulations, with random positions (RA and Dec) in sky allocated to AGNs in our sample and then a mock dipole was superimposed to calculate Doppler boosting and aberration effects for each source. Then on this mock catalogue of MIRAGNs, our 3-d COSFIT procedure was applied to recover the dipole and compared with the input dipole in that simulation. This not only validated our method but it also provided us an estimate of errors from 500 independent simulations. The correction factor was determined from another set of 500 Monte-Carlo simulations, run without a gap about the galactic plane, to be $k_2\approx 1.15$ in this method.

Accordingly, for the direction of the peculiar velocity, along with estimated errors, we arrive at RA$=148^{\circ}\pm 19^{\circ}$, Dec$=22^{\circ}\pm 18^{\circ}$ from the the Chi-square ($\chi^2_\nu$) fit which shows a minimum (Figure~7b) at the same sky position as the peak in the dipole magnitude (Figure~7a). The $\chi^2_\nu$ minimum value of 1.0, which is an ideal value, indicates that the estimates of errors in the dipole amplitudes are quite realistic. The value for the MIRAGN dipole direction agrees, within the $2\sigma$ uncertainty, with the CMBR dipole direction, RA$=168^{\circ}$, Dec$=-7^{\circ}$. However, the peculiar speed, determined from, $v\approx c D /(5.4 k_2)\approx4.8 \times 10^4 {\cal D}$ km s$^{-1}$, gives  a solar speed value $1.6\pm 0.2 \times 10^{3}$ km s$^{-1}$, $\stackrel{>}{_{\sim}} 4$ times the CMBR value.

\begin{figure*}
\centering
\includegraphics[width=\linewidth]{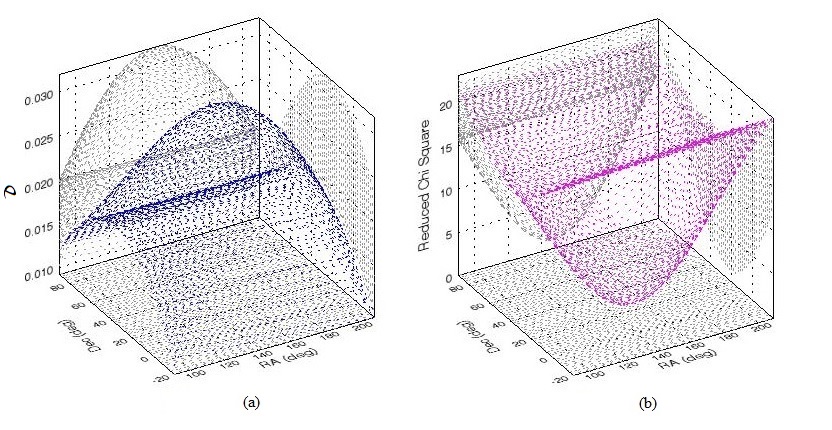}
\caption{A plot of 3-d  COSFITs made to the dipole amplitudes estimated for various trial dipole directions across the sky, showing (a) a unique peak (blue colour) unambiguously indicating the optimum direction of the dipole (b) reduced chi-square ($\chi^2_\nu$) values (magenta colour), having an ideal minimum value of 1.0, at the same sky position as the blue colour peak in (a). The horizontal axes denote Right Ascension (RA) and Declination (Dec) in degrees. 
The positions (RA and Dec) of the extrema are determined more easily from the 2-d projections, shown in light grey, thence we infer the direction of the observer's peculiar velocity as RA$=148^{\circ}$, and Dec$=22^{\circ}$.} 
\end{figure*}

Our results are presented in Table 3, where the 1st column gives the magnitude range, 2nd column gives the number of sources, 3rd column gives the difference (${N}1-{N}2$) between the numbers of sources in ${\cal S}_1$ and ${\cal S}_2$, 4th and 5th columns give the direction of the dipole in terms of Right Ascension and Declination, 6th column gives $\cal D$, the dipole magnitude, and the last column gives the value of the speed estimated from $\cal D$.
Dipole ${D}$ was estimated for samples 
containing all sources in our sample as well as for the three bins with different magnitude (W1) limits. The direction (RA and Dec) of the dipole in each case was determined from a 3-d COSFIT and the velocity $v$ then determined, taking the pole to be in that direction.

From Tables~1 and 3 we notice that the uncertainty in the peculiar velocity values ($0.2 - 0.4 \times 10^{3}$ km s$^{-1}$) determined from the MIRAGN data is of the same order as the CMBR dipole amplitude ($ 0.37 \times 10^{3}$ km s$^{-1}$), from that it is evident that a positive detection of the  MIRAGN dipole has been possible only because the strength of the signal has turned out to be much larger ($>4$) than the CMBR dipole. It has been pointed out that due to statistical fluctuations, there could be a bias in the dipole amplitudes towards higher values \cite{5}, and which, even in the absence of a genuine dipole could yield a value approximately equal to the CMBR dipole value (of course in a random direction), or for a dipole actually equal to the CMBR value, could lead to an estimate of the dipole magnitude substantially higher (by a factor $\sim 1.4$). However, at the level of much larger dipole amplitudes, like our observed ${\cal D} \sim 3 \times 10^{-2}$ (Tables 1 and 3), the effects of the amplitude bias are at most $\stackrel{<}{_{\sim}} 3\%$, a negligible fraction of the quoted uncertainties in the Tables. 
\section{Discussion}
The view that the Solar system peculiar motion is 370 km s$^{-1}$, as given by the CMBR dipole, seems to be widely accepted. However, one needs to keep an open mind about it, without sticking rigidly to the conventional view that only the CMBR provides a reference frame which, somehow, is to be considered as more fundamental than the other ones, e.g. AGNs, for establishing the peculiar motion of the solar system. While the CMBR refers to the radiation era ($z\sim 10^3$), AGNs represent the much later matter era ($z\sim 1-3$). 
A common direction for the two dipoles, determined from completely independent data, does indicate that these dipole amplitudes differ not because of random statistical fluctuations, or due to some systematics in the observations or in the data analysis, otherwise even their directions would have turned out to be different, and that the difference in their amplitudes might be a pointer to some occurrence of cosmological significance.

In Figure 3, we saw that there are large scale inhomogeneities in the sky distribution of the MIRAGN number densities at W1$>16$ levels. In fact, even at somewhat brighter levels, $15.5<$W1$<16$, such large scale inhomogeneities in the sky distribution are present (Figure 2), though at a reduced level. How can we be sure that such large scale inhomogeneities in the sky distribution are not present in our sample at W1$<15$ levels? Although we may not be able to discern them from Figure 1 with our eyes, but how to make sure of the lack of their underlying presence in our data, at least at levels which could have influenced our intended determination of the dipole significantly. 
In other words, could such large scale inhomogeneities or even some large scale statistical fluctuations in certain directions be the cause of our observed dipole being much larger than the CMBR dipole?
\begin{table*}[t]
\begin{center}
\caption{The dipole magnitude estimated using differential number counts in the hemisphere method}
\hskip4pc\vbox{\columnwidth=33pc
\begin{tabular}{ccccccccccccccccc}
 \hline
(1)&&(2)&(3)&(4)&(5)&(6)&(7)\\ 
Magnitude Range& &$N$  &$N_1-N_2$& RA & Dec  &     $\cal D$  & $v$ \\
 &&&& ($^{\circ}$)&  ($^{\circ}$)&($ 10^{-2}$) &   ($10^{3}$ km s$^{-1}$)
\\  
\hline
$15.0>W1\geq 12.0$  &  & 279139 &  4629& $148\pm 19$ & $22\pm 18$  & $3.3\pm 0.4$ &  1.6 $\pm 0.2$ \\
$15.0>W1\geq 14.7$ &  & 102822 &  2370& $157\pm 20$ & $22\pm 19$  & $4.6\pm 0.6$ &   2.2 $\pm 0.3$  \\
$14.7>W1\geq 14.3$ &  & 086035 &  1211& $132\pm 20$ & $30\pm 19$  & $2.9\pm 0.7$ &   1.4 $\pm 0.4$  \\
$14.3>W1\geq 12.0$  &  & 090282 &  1022& $143\pm20 $ & $10\pm 19$  & $2.1\pm 0.7$ &   1.0 $\pm 0.4$  \\
 \hline
\end{tabular}
}
\end{center}
\end{table*}

\begin{figure*}[t]
\centering
\includegraphics[width=\linewidth]{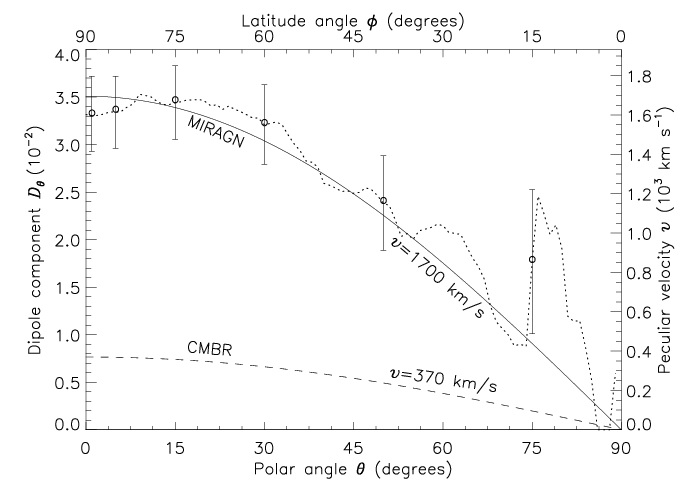}
\caption{A plot of the dipole components ${\cal D}_{\theta}$ and the equivalent peculiar velocity component $v \cos\theta$, computed for various zones of the sky between the great circle and a parallel circle at $\theta$, the angle with respect to the determined dipole direction, RA$=148^{\circ}$, Dec$=22^{\circ}$. The latitude angle $\phi$ is measured from the great circle at $\theta=90^\circ$. The observed values in  various sky strips are plotted as circles (o), with the error bars calculated for a random (binomial) distribution. 
The continuous curve, corresponding to a peculiar velocity of  $1.7\pm 0.2 \times 10^{3}$ km s$^{-1}$, shows the expected ($\propto \cos \theta$) behaviour for ${\cal D}_{\theta}$, which is a best fit to the data observed in different sky strips, shown by the dotted curve. For a comparison, the expectation for the CMBR value, $v=370$ km s$^{-1}$, is shown by dashed curve, which lies way below the values observed for the MIRAGNs.}
\end{figure*}

\begin{figure}[t]
\includegraphics[width=\linewidth]{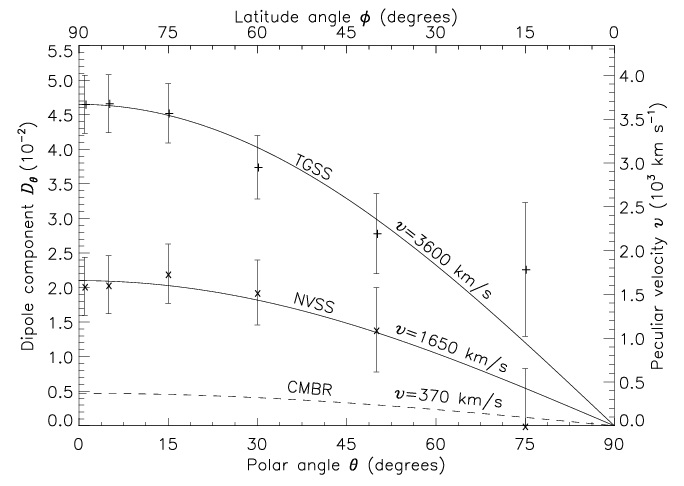}
\caption{Plots of the dipole components ${\cal D}_{\theta}$, observed for the TGSS (+) and  NVSS (x) data \cite{9} for various zones of the sky between the great circle and a parallel circle at $\theta$, the angle with respect to the CMBR dipole direction. The uncertainties expected in random (binomial) distributions are shown as error bars. 
The continuous curves, corresponding to the indicated peculiar velocities, show the expected ($ \cos \theta$) behaviour for ${\cal D}_{\theta}$, which seem to fit well the observed values. The dashed curve shows the plot, expected for the CMBR value, $v=370$ km s$^{-1}$.}
\end{figure}

For one thing, such non-uniformities would have to be distributed over various sky regions in such a way that these make the resulting excess in the dipole amplitude to occur in the same direction as the CMBR dipole. Moreover, these also have to be spread all across the magnitude band of our MIRAGN sample, as our three bins, selected with no overlap in the magnitude band, yield very similar dipole vectors.  In fact, one can verify that it is a genuine dipole distribution in the number densities and not merely a result of some systematics in the data. From Eq.~(\ref{eq:8}), we see that ${\cal D}_{\theta}$ yields a component of the dipole along $\theta$. We can verify this  $\cos\theta$ dependence of ${\cal D}_{\theta}$, by making cumulative counts of $N_1(\theta)$ and $N_2(\theta)$ in two opposite hemispheres, ${\cal S}_1$ and ${\cal S}_2$, for various $\theta$ values. Figure~8 shows a plot of the ${\cal D}_{\theta}$, computed for our MIRAGN sample using Eq.~(\ref{eq:8}), as a function of $\theta$, the angle with respect to the determined dipole direction, RA$=148^{\circ}$, Dec$=22^{\circ}$. The dotted curve shows ${\cal D}_{\theta}$ calculated from cumulative counts which has the best fit by a $v \cos\theta$ dipole curve, with $v=1.7\pm 0.2 \times 10^{3}$ km s$^{-1}$. We have also plotted the error bars, calculated for a random (binomial) distribution, at some representative points.
The larger fluctuations seen for ${\theta}\rightarrow 90^\circ$, although still within statistical uncertainties, are because of a smaller dipole component (${\cal D}_{\theta}\propto\cos\theta$) vis-\`a-vis the relatively larger statistical fluctuations in a binomial distribution due to a lesser number of sources in the reduced sky zones. 
Also shown is the expected curve for the peculiar velocity, $v=370$ km s$^{-1}$ derived from the CMBR dipole, which lies way below the dipole components estimated from the MIRAGNs data.

A vectorial summation of sky positions of all sources, as determined in Table 1, in itself does not guarantee that the resultant vector is necessarily a dipole, after all we will always get some resulting RA, Dec and $\cal D$ from such a summation, whatever might be the statistical uncertainties. However, a systematic variation pattern of dipole amplitudes over sky, as seen in Figure~5, where the pole position $M$ from Table 1 is right in the middle of the darkest region, while the corresponding antipole position $A$ lies in the lighter most region, indicates the genuine nature of the dipole. 
Moreover, the dashed curve, representing the zero amplitude of the dipole, overlaps the great circle, at $90^\circ$ from the poles at $M$ and $A$, added to
the fact that we obtained the same particulars for the dipole from the hemisphere method, where not only the direction of the dipole from the brute force method (Table 3) turned out to be the same as was determined in Table 1, even the amplitude of the dipole, where various sources, depending upon their sky positions get different weights, gave the same value, confirms that it is a genuine dipole. More so because of the  $\cos\theta$ pattern seen in Figure~8, as would be expected only for a dipole.
Only in a rather contrived scenario would one expect the differential number counts in two opposite hemispheres in sky, to follow the $\cos\theta$ behaviour, unless it were the result of a genuine dipole in the number density distribution, irrespective of the ultimate cause of the dipole. 

Though spectroscopic redshifts are not available for most of the sources in the sample, however, from the sources where redshifts are available \cite{17}, as well as from the previous information in the literature, AGNs are known to lie mostly at redshifts between 0.5 and 3 and are, in fact, the most distant discrete objects seen in the universe. Therefore, the dipole anisotropy seen among them is not a local effect, and has a genuine bearing on the cosmological principle. 

For a comparison with some earlier dipole determinations from the radio data on AGNs \cite{9}, in Figure~9 we show the plots of the ${\cal D}_{\theta}$ for both TGSS and NVSS data as a function of $\theta$, measured with respect to the CMBR dipole direction.  It should be noted that the conversion factor from $\cal D$ to peculiar velocity $v$ in Figure~9 is somewhat different in comparison with that in Figure (7), because of the difference in indices $x$ and $\alpha$ between the radio and Infra red populations.
It does seem that the peculiar velocity of the Solar system estimated from the TGSS data is much higher than the NVSS data \cite{9}, the latter in agreement with the MIRAGN dipole in Figure~8. 

\begin{figure*}[h]
\centering
\includegraphics[width=12cm]{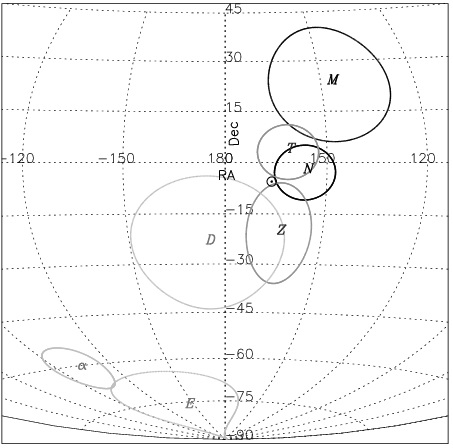}
\caption{The sky position, in equatorial coordinates RA and Dec, of the pole determined from our MIRAGN sample is indicated by $M$, along with the error ellipse, while other pole positions for various dipoles shown on the map are: $N$ (NVSS) \cite{4}, $T$ (TGSS) \cite{9}, $Z$ (DR12Q) \cite{10}, $D$ (Dark Flow) \cite{Ka10}, $\alpha$ (Fine Structure constant) \cite{Ki12,Ma16} , and $E$ (Dark Energy) \cite{Ma12}. The CMBR pole, indicated by $\odot$, has negligible errors \cite{3}.}
\end{figure*}
A recent determination of the peculiar motion of the Solar system from a dipole anisotropy in the redshift distribution of distant quasars has also yielded discrepant value of 
$-2350\pm 280$ km s$^{-1}$ along the direction of the CMBR dipole \cite{10}, implying a solar peculiar motion in a direction opposite to that derived from all other previous dipoles.
Nevertheless, it is evident that all AGN dipoles have much larger amplitudes than the CMBR dipole, even though various AGN dipole directions in sky may be lying parallel to the CMBR dipole.

In Figure~10, the sky position of our MIRAGN  pole is shown by M, along with the $1\sigma$ error ellipse, while the CMBR pole position is indicated by $\odot$, which has negligible errors \cite{1,2,3}. Also plotted are the various other determined pole positions, N for the NVSS dipole, T for the TGSS dipole and D for the DR12Q dipole, along with their $1\sigma$ error ellipses \cite{4,9,10}. From Figure 10, it does seem that various AGN dipoles are within $\sim 1\sigma$, except for the MIRAGN pole which is at $\stackrel{<}{_{\sim}} 2\sigma$, of the CMBR pole, from which we can surmise that the various AGN dipoles are pointing along the same direction as the CMBR dipole.

Here it may be pointed out that a recent determination of the dipole \cite{11} from a flux-limited, all-sky sample of 1.36 million quasars observed by the Wide-field Infrared Survey Explorer (WISE) gave the direction of the dipole to be similar, $l=238\!\stackrel{^\circ}{_\cdot}\!2, b=28\!\stackrel{^\circ}{_\cdot}\!8$ (RA=$140\!\stackrel{^\circ}{_\cdot}\!0$, Dec=$-6\!\stackrel{^\circ}{_\cdot}\!5)$, about $27\!\stackrel{^\circ}{_\cdot}\!8$ from the CMBR pole. However, the amplitude of the peculiar velocity ($8.2\times 10^2$ km s$^{-1}$) was over twice as large the CMBR value ($3.7\times 10^2$ km s$^{-1}$), while we find the peculiar velocity ($1.7\times 10^3$ km s$^{-1}$) to be over four times the CMBR value. This mysterious factor of two difference 
in the peculiar velocity values, estimated from these two Mid Infra Red AGN samples, needs to be explored further, whether it is due to a difference in the basic samples themselves or is it arising from the differences in the excluded zones, e.g. the galactic latitudes, or caused by some other, as yet unknown, reason. Nevertheless, one thing is clear, both the peculiar velocity  estimates are inconsistent with the CMBR peculiar velocity value, though the direction estimate in each individual case does appear consistent with the CMBR dipole direction, within statistical uncertainties.
 
Since the peculiar velocity of the Solar system should not dependent upon the specific data or technique used to determine it, one obvious inference from the discordant values of the inferred peculiar motion from observed dipoles for the CMBR and the AGNs could be that there is a large relative motion between various cosmic reference frames. Otherwise, one has to take the view point that the observed discordant dipoles (including the CMBR), contrary to the conventional wisdom, do not reflect a motion of the observer (or a peculiar velocity of the Solar system), and that we should instead look for some other possible cause for the genesis of these dipoles, including that of the CMBR. However, one has to then explain the existence of a preferred  direction of all the dipoles and that what is so special about this particular direction 
A not-too-far-fetched inference drawn could be that a common direction for the dipoles, including of the CMBR, is a pointer toward the presence of an inherently preferred cosmic direction (axis!), implying perhaps an anisotropic Universe \cite{4} in conflict with the Cosmological Principle. 

There are also other dipoles seen at cosmological scales. 
For instance, the ``dark flow'' dipole, which is a statistically significant dipole found at the position of galaxy clusters in filtered maps of the CMBR temperature anisotropies, implies the existence of a primordial CMBR dipole of non-kinematic origin, presenting itself
as an effective motion across the entire cosmological horizon \cite{Ka08,Ka09,Ka10,At15}. 
The CMBR dipole measurements are also consistent with some dynamical contributions to the CMBR dipole, resulting possibly from an SU(2) gauge principle \cite{Sz08,Lu09,Ho13}. The detection of a dipole in the spatial variation of the fine-structure constant, $\alpha = e^2/\hbar c$, based on study of quasar absorption systems has also been reported \cite{We11,Ki12,Be11,Be12}. These spatial variations of the fine structure constant can be constrained using clusters of galaxies \cite{Ga13,Ma16}.
It seems that the fine structure constant cosmic dipole is aligned with the corresponding dark energy dipole within 1$\sigma$ uncertainties  \cite{Ma12,Ma13}.
In Figure 10, we have plotted positions of some of these poles to show their correlation, if any, with the CMBR dipole as well as various AGN dipoles. Here $D$ (Dark Flow) \cite{Ka10}, $\alpha$ (Fine Structure constant) \cite{Ki12,Ma16} , and $E$ (Dark Energy) \cite{Ma12} are shown along with their quoted error ellipses. While 
the direction of dark flow dipole could be correlating with the  direction of the CMBR dipole, the fine structure constant cosmic dipole and the dark energy dipole seem to be pointing quite away from the direction of the CMBR dipole. At the same time, because of large statistical  uncertainties. the direction of dark flow dipole being in the same direction as the dark energy dipole cannot be ruled out.
Further, there is a preference seen at significant levels for odd multipoles in the large scale  anisotropies of the CMB temperature, a parity asymmetry, to be strongly aligned with the CMBR dipole \cite{Na12,Zh14,Ch16}. 

If the CMBR dipole is of kinematic origin due to the Solar system peculiar motion, something resulting from the distribution of matter at local scales, in contrast to dipoles originating at cosmic scales,  these observed correlations in the directions of various dipoles and asymmetries give rise to doubts about whether the CMBR dipole is really caused by our peculiar motion. And if the CMBR dipole is due to some intrinsic asymmetry at cosmic scales, suggested by an order of magnitude difference in the dipole magnitudes as inferred from various AGN datasets, even though all appear to be in the same direction as the CMBR dipole, it gives rise to a larger question of the ultimate reliability of the Cosmological Principle, a cornerstone of the modern cosmology.
\section{Conclusions}
From the angular positions in sky of a sample of ~0.28 million Mid Infra Red AGNs, we found an anisotropy in their number densities in different directions. Ascribing this anisotropy to the peculiar motion of the observer, we determined the peculiar velocity of the Solar system that turned out to be, like other earlier AGN dipoles, at least a factor of four larger that that inferred from the CMBR dipole, but along the same direction. Since the peculiar velocity of the Solar system should not depend upon the specific data or the technique used to determine it, a question gets raised about the nature of these dipoles seen in the sky and whether the genesis of some or all of these dipoles indeed is due to the peculiar motion of the Solar system. A common direction for all these dipoles, including the CMBR one, determined from completely independent surveys by different groups, does indicate that the differences in the dipole amplitudes are genuine and not because of random statistical fluctuations, or due to some systematics in the observations or in the data analysis, otherwise even the dipole directions obtained from different data sets would have been different. An inference that could possibly be drawn from a common direction for all the dipoles is that it might be a pointer toward the presence of an inherently preferred cosmic direction (axis!), implying perhaps an anisotropic Universe, in discordance with the Cosmological Principle, a cornerstone of the modern cosmology. 

\vspace{6pt}
\funding{No funds, grants, or other support of any kind was received from anywhere for this research.}
\conflictsofinterest{The author has no conflicts of interest to declare that are relevant to the content of this paper.} 
\reftitle{References}

\end{document}